\documentclass[reprint,nofootinbib,superscriptaddress,amsmath,amssymb,aps,prd]{revtex4-1}
\usepackage{graphicx}
\usepackage{rotating}
\usepackage{dcolumn}
\usepackage{bm}
\usepackage{color}
\usepackage{mathptmx, textcomp}
\usepackage[latin1]{inputenc}
\usepackage{braket}
\usepackage[colorlinks,linkcolor=magenta,anchorcolor=cyan,citecolor=blue]{hyperref}
\usepackage[multiple]{footmisc}

\def\be{\begin{equation}}
\def\ee{\end{equation}}
\def\ba{\begin{eqnarray}}
\def\ea{\end{eqnarray}}

\newcommand{\eq}[1]{(\ref{#1})}

 \def\w{\omega}\def\r {\rho}        \def\d {\delta} \def\f {\phi}     \def\l {\lambda} \def\z {\zeta} \def\x {\xi} \def\c {\chi}   \def\m {\mu} \def\pd {\partial} \def \inf {\infty}  
\def\Q{\Theta} \def\W{\Omega}     \def\S {\Sigma} \def\D {\Delta}       \def\grad{\nabla}\def\.{\cdot}
\def\math {\mathcal}
\begin{document}

\title{Universality of entropy principle for a general diffeomorphism-covariant purely gravitational theory}
\author{Jie Jiang}
\email{jiejiang@mail.bnu.edu.cn}
\affiliation{Department of Physics, Key Laboratory of Low Dimensional Quantum Structures and Quantum Control of Ministry of Education, and Synergetic Innovation Center for Quantum Effects and Applications, Hunan Normal University, Changsha 410081, Hunan, People's Republic of China}
\affiliation{
College of Education for the Future, Beijing Normal University, Zhuhai 519087, China}
\affiliation{Department of Physics, Beijing Normal University, Beijing, 100875, China}
\author{Xiongjun Fang}
\email{Corresponding author: fangxj@hunnu.edu.cn}
\affiliation{Department of Physics, Key Laboratory of Low Dimensional Quantum Structures and Quantum Control of Ministry of Education, and Synergetic Innovation Center for Quantum Effects and Applications, Hunan Normal University, Changsha 410081, Hunan, People's Republic of China}
\author{Sijie Gao}
\email{sijie@bnu.edu.cn}
\affiliation{Department of Physics, Beijing Normal University, Beijing, 100875, China}

\begin{abstract}
Thermodynamics plays an important role in gravitational theories. It is a principle independent of the gravitational dynamics, and there is still no rigorous proof to show that it is consistent with the dynamical principle. We consider a self-gravitating perfect fluid system in a general diffeomorphism-covariant purely gravitational theory. Based on the Noether charge method proposed by Iyer and Wald, considering static off/on-shell variational configurations which satisfy the gravitational constraint equation, we rigorously prove that the extrema of the total entropy of perfect fluid inside a compact region for fixed total particle number demands that the static configuration is an on-shell solution  after we introduce some appropriate boundary conditions, i.e., it also satisfies the spatial gravitational equations. This means that the entropy principle of the fluid stores the same information as the gravitational equation in a static configuration. Our proof is universal and holds for any diffeomorphism-covariant purely gravitational theories, such as Einstein gravity, $f(R)$ gravity, Lovelock gravity, $f($Gauss-Bonnet$)$ gravity and Einstein-Weyl gravity. Our result shows the consistency between the ordinary thermodynamics and the gravitational dynamics.

\end{abstract}
\maketitle
\section{Introduction}

Black holes are important and fundamental objects in the gravitational theories after the general relativity was proposed.
{ In the past few decades, many researches in general relativity has implied that black holes can be regarded as a thermodynamical system. The four laws of black hole mechanics in general relativity has been constructed in Refs. \cite{A2, A3, A31}. The discovery of the Hawking radiation provided a natural interpretation of the laws of the black hole mechanics as the ordinary laws of thermodynamics \cite{A4}.} Since then, black hole thermodynamics has attracted lots of attention, and people believe that it could provide us with a deeper understanding of the gravitational theories.

It is generally believed that the gravitational equation is the basic equation of nature. Therefore, as a traditional way, people always study how to construct the laws of thermodynamics from a gravitational dynamics. For instance, Wald generally derived the first law of the black hole thermodynamics for general diffeomorphism-covariant gravitational theory based on the Noether charge method \cite{Wald01,IW}, and Wall also discussed the second law for general higher curvature gravity \cite{Wall}. However, some people believe that the thermodynamical relations are the more fundamental assumption, and then the gravitational equations should be derived from thermodynamics. From this point of view, Jacobson considered that the gravitational equations are the equation of state and the Einstein equation can be derived from the first thermodynamic law on the local Rindler horizons \cite{Jaco}. After that, this idea has been accepted by more and more people in the past years \cite{CaiKim,Verlinde,Brabetti}. { All these discussions showed the consistency between the gravitational thermodynamics and the gravitational dynamics.}


{ In a spacetime without event horizon, there also exists some matter field, such as self-gravitating perfect fluid, which satisfies the ordinary thermodynamical laws.} In contrast to the black hole thermodynamics, the local thermodynamical quantities of the fluid, such as entropy density $s$, energy density $\r$, and local temperature $T$, are well defined. { From the viewpoint of the gravitational dynamics, the on-shell static distribution of these local quantities for perfect fluid can be determined by the gravitational equations. Moreover, from the viewpoint of the thermodynamics, since the static self-gravitating fluid can be regarded as a thermodynamical system, the on-shell distribution can also be obtained by extrema of the total entropy of the perfect fluid in the off/on-shell static variation. In general, the thermodynamics of the fluid in the spacetime and the gravitational dynamics are two independent principles. However, if we believe that both of them are reliable, they should give the same distribution of the fluid.} Therefore, Cocke proposed the entropy principle for the self-gravitating fluid. It states that under a few natural conditions, the extrema of total entropy of perfect fluid are equivalent to the Einstein equation in the static self-gravitating fluid system \cite{Cocke}. For the spherical radiation system, Sorkin, Wald, and Zhang have shown that the Tolman-Oppenheimer-Volkoff equation can be derived from the Einstein constraint equation (time-time component of the gravitational equations) by the extrema of the total entropy of perfect fluid inside a compact region \cite{SWZ}. Gao extended their work to an arbitrary perfect fluid in a static spherical geometry \cite{Gao}, { in which they only used some thermodynamical relations.} After that, this principle has been widely studied by the researchers \cite{GSW,Roupas,Lovelock1,B1,B2,B3}. { Their results showed the consistency between the ordinary thermodynamics of fluid and gravitational dynamics.}

Most recently, the entropy principle is generally proved in the static spacetime without the spherical symmetry for Einstein-Maxwell gravity \cite{GR,EM}, Lovelock gravity \cite{Lovelock2}, and $f(R)$ gravity \cite{Fang:2015pcw}. Although all of them showed the validity of the entropy principle, there is still a lack of a general proof to show that it is valid for any gravitational theories. When the string effects or quantum corrections are taken into account, the effective gravitational action should be corrected by the powers of the curvature tensor and its derivatives, and the Einstein-Hilbert action is just the first term of the effective action \cite{H1,H2,H3,H4,H5,H6}. { Moreover, there is some literature that modified the Einstein gravity by introducing the higher-curvature corrections,} such as $f(R)$ term and $f($Gauss-Bonnet$)$ term, to deal with the problems in cosmology and astrophysics \cite{H7,H8,H9}. Thus, it is natural for us to ask whether the entropy principle is satisfied for these modified  gravitational theories. We can see that all of these modified theories can be described by a diffeomorphism-invariant action. Therefore, in the following, we would like to prove the equivalence of the extrema of the total entropy of perfect fluid inside a compact region in the off/on-shell static configurations and the gravitational equations for a general diffeomorphism-covariant { purely} gravitational theory in which the Lagrangian is constructed by metric and its derivatives only. To prove the equivalence, we first present a theorem related to the entropy principle:

\textbf{Theorem}: Consider a variation related to a one-parameter family of the off/on-shell static configurations which satisfy the gravitational constraint equation (the off-shell configuration doesn't obey the spatial gravitational equation) in a $n$-dimensional spacetime for a general diffeomorphism-covariant purely gravitational theory coupled to a self-gravitating perfect fluid. Denote the $(n-1)$-dimensional hypersurface $\S$ to be a moment of the static observers. Choose $C$ to be a compact region inside the hypersurface $\S$ with a boundary $\pd C$. Assume that the fluid velocity coincides with the static Killing vector and the local temperature $T$ of the fluid obeys the Tolman's law. After introducing the appropriate boundary conditions which keep the quasi-local conserved charge of the static Killing vector inside $C$ fix, the extrema of the total entropy of perfect fluid inside $C$ for fixed total particle number in this off/on-shell variation demands that the static configuration is an on-shell solution (i.e., it also satisfies the spatial gravitational equations).

Our paper is organized as follows. In the next section, we discuss dynamical and thermodynamical features of the off/on-shell static configuration for a general diffeomorphism-covariant gravitational theory. In Sec. \ref{sec3}, we prove the \textbf{Theorem} of the entropy principle based on the Noether charge method proposed by Iyer and Wald \cite{IW}. Finally, the conclusions are presented in Sec. \ref{conc}.

\section{Static configuration of a self-gravitating perfect fluid}\label{sec2}

In this section, we start by discussing the properties of the static configuration in a general $n$-dimensional diffeomorphism-covariant { purely} gravitational theory sourced by a self-gravitating perfect fluid. As mentioned above, we would like to consider a static configuration in $n$-dimensional spacetime. In this situation, there exists a static Killing vector field $\x^a$ satisfying $\math{L}_\x g_{ab}=\grad_{(a} \x_{b)}=0$. The integral curves of $\x^a$ are the worldlines of the static observers in the spacetime. The velocity vector field is given by $u^a=\c^{-1} \x^a$, in which $\c=\sqrt{-\x^a\x_a}$ is the red-shift factor for the static observers and $u^a$ is also the normal vector on $\S$. The induced metric $h^{ab}$ on $\S$ is given by
\ba\begin{aligned}
h^{ab}=g^{ab}+\c^{-2}\x^a\x^b\,.
\end{aligned}\ea
The Lagrangian $n$-form in this theory is given by
\ba\begin{aligned}
\bm{L}=\bm{L}_\text{grav}+\bm{L}_\text{fluid}\,,
\end{aligned}\ea
in which $\bm{L}_\text{grav}$ and $\bm{L}_\text{mt}$ are the gravitational part and fluid part of the Lagrangian, separately. The gravitational part of the Lagrangian is a function of the metric $g_{ab}$, Riemann tensor $R_{bcde}$ and its higher-order derivative $\grad_{a_1}\cdots \grad_{a_k} R_{bcde}$. Using the relationship
\ba
2\grad_{[a}\grad_{b]}T_{c_1\cdots c_k}=\sum_{i=1}^{i=k}R_{abc_i}{}^d T_{c_1\cdots d \cdots c_k}
\ea
to exchange the indices of the derivative operators, it is not hard to verify that the Lagrangian $\bm{L}$ can be reexpressed as
\ba\begin{aligned}
\bm{L}_\text{grav}=\bm{\epsilon}\math{L}_\text{grav}(g_{ab}, R_{bcde}, \cdots, \grad_{(a_1}\cdots \grad_{a_k)}R_{bcde},\cdots)\,,
\end{aligned}\ea
This is exactly the expression of the Lagrangian considered in \cite{IW} for a general diffeomorphism-covariant gravitational theory. In the discussion, we use boldface symbols to denote the differential { forms} in the spacetime. The gravitational equations are given by the variation of the Lagrangian and we can express them by \cite{IW}
\ba\begin{aligned}
\math{E}_{ab}=H_{ab}-T_{ab}
\end{aligned}\ea
with
\ba\begin{aligned}
H_{ab}=A_{ab}+P_{acde}R_b{}^{cde}-2\grad^c\grad^d P_{acdb}-\frac{1}{2} g_{ab}\math{L}_\text{grav}
\end{aligned}\ea
and the stress-energy tensor of the perfect fluid
\ba\begin{aligned}
T_{ab}=\r u_a u_b+p (g_{ab}+u_a u_b)\,,
\end{aligned}\ea
in which $\r$ and $p$ are the energy density and pressure of the fluid separately, and we have denoted
\ba\begin{aligned}
A_{ab}=\frac{\pd \math{L}_\text{grav}}{\pd g^{ab}}\,,\quad P^{abcd}=\frac{\d \math{L}_\text{grav}}{\d R_{abcd}}\,.
\end{aligned}\ea
The expression of the stress-energy tensor implies that the static observers are also the comoving observers of the fluid.

The \textbf{Theorem} of the entropy principle shows that we need to consider an off/on-shell static configuration which satisfies the gravitational constraint equation (time-time component of the equation of motion)
\ba\begin{aligned}\label{constraint}
\r= H_{uu}=H_{ab}u^a u^b\,.
\end{aligned}\ea
Therefore, the ``off-shell'' refers specifically to the off-shell static configuration which does not obey the spatial gravitational equations
\ba\label{spatialeq}
p h_{ab}\neq\hat{H}_{ab}=h_{ac}h_{bd}H^{cd}\,.
\ea

In the following, we consider the thermodynamics of the self-gravitating perfect fluid which satisfies the Tolman's law $T\c=T_0$ in a static configuration, where $T_0$ is a constant and it can be regarded as the red-shift temperature. Without loss of generality, we shall take $T_0=1$ such that
\ba\begin{aligned}
T=\c^{-1}\,.
\end{aligned}\ea
The entropy density $s$ is a function of the energy density $\r$ and the particle number density $n$, i.e., $s=s(\r ,n)$. From the familiar first law on the region $C$, one can derive the local first law and the Gibbs-Duhem relation of the fluid \cite{Gao},
\ba\begin{aligned}\label{firstlaw}
d\r=Tds+\m dn\,,\quad\quad \r=T s-p+\m n\,,
\end{aligned}\ea
where $\m$ is the chemical potential corresponding to the particle number density $n$. From the local first law, we can see that the entropy of perfect fluid can be treated as a function of the energy density $\r$ and particle number density $n$, i.e., $s=s(\r, n)$. The conservation law $\grad_a T^{ab}=0$ for the perfect fluid gives
\ba\begin{aligned}
dp+\c^{-1}(\r+p)d\c=0\,.
\end{aligned}\ea
Together with the local first law and Gibbs-Duhem relation in Eq. \eq{firstlaw}, we can further obtain the result that
\ba\begin{aligned}
\m \c=\text{constant}\,.
\end{aligned}\ea

\section{Noether charge method and the proof of the entropy principle}\label{sec3}

In this section, we would like to { prove} the entropy principle based on the Neother charge method proposed by Iyer and Wald \cite{IW}. We consider a one-family $\f(\l)$ of the off/on-shell static field configurations as described in the last section, in which we denote $\f(\l)$ to metric $g_{ab}(\l)$ and the shelf-gravitating perfect fluid with $\r(\l), p(\l)$. That is to say, $\f(\l)$ satisfy the thermodynamical properties of the fluid as described in the last section as well as the gravitational constraint equation
\ba\begin{aligned}\label{rH}
H_{uu}(\l)=\r(\l)\,.
\end{aligned}\ea
For the off-shell configuration $\f(\l)$, the spatial gravitational equations are not satisfied, i.e.,
\ba\begin{aligned}
\hat{H}_{ab}(\l)\neq p(\l) h_{ab}(\l)\,.
\end{aligned}\ea
In the following, we will define the notations
\ba\begin{aligned}
\c=\c(0)\,,\quad\quad\d\c=\left.\frac{\pd \c}{\pd\l}\right|_{\l=0}
\end{aligned}\ea
to denote the background quantity and its variation in the family $\f(\l)$.
Considering the diffeomorphism covariance of the theory, we can choose a gauge to fix the static Killing vector $\x^a(\l)$ under the variation in the static configuration $\f(\l)$, i.e., $\d \x^a=0$. For each static configuration $\f(\l)$, we have
\ba\begin{aligned}
g^{ab}(\l)=-\c(\l)^{-2}\x^a\x^b+h^{ab}(\l)\,.
\end{aligned}\ea
Then, we have
\ba\begin{aligned}\label{decdg}
\d g^{ab}&=2\c^{-3}\d\c\x^a\x^b+\d h^{ab}\,.
\end{aligned}\ea

In this family, the variation of the gravitational part of the Lagrangian gives
\ba\begin{aligned}
\d \bm{L}_\text{grav}=\bm{E}_{ab}^\text{grav}\d g^{ab}+d\bm{\Q}^\text{grav}(g, \d g)\,,
\end{aligned}\ea
in which
\ba\label{EH}
\bm{E}_{ab}^\text{grav}=\frac{1}{2}\bm{\epsilon} H_{ab}
\ea
denotes the gravitational part of the equation of motion, and $\bm{\Q}^\text{grav}(g, \d g)$ is the symplectic potential. After completing all the indices, Eq. \eq{EH} is expressed as $(E_{ab}^\text{grav})_{a_1\cdots a_n}=\epsilon_{a_1\cdots a_n}H_{ab}$.

For any vector field $\z^a$, we can define a Noether current $(n-1)$-form as
\ba\begin{aligned}\label{J1}
\bm{J}_\z^\text{grav}=\bm{\Q}^\text{grav}(g,\math{L}_\z g)-\z\.\bm{L}_\text{grav}\,.
\end{aligned}\ea
It has been shown in \cite{IW} that it can be expressed as
\ba\begin{aligned}\label{J2}
\bm{J}_\z^\text{grav}=\bm{C}_\z^\text{grav}+d\bm{Q}_\z^\text{grav}\,,
\end{aligned}\ea
in which $C_\z^\text{grav}=\z\.\bm{C}^\text{grav}$ with
\ba\begin{aligned}\label{ExpC}
\bm{C}^\text{grav}_{aa_1\cdots a_{n-1}}=\bm{\epsilon}_{ba_1\cdots a_{n-1}}H_a{}^b
\end{aligned}\ea
is the constraint of the gravitational theory and $\bm{Q}_\z^\text{grav}$ is a Neother charge $(n-2)$-form of the vector field $\z^a$.

Using the two expressions \eq{J1} and \eq{J2} of the Noether current, we have
\ba\begin{aligned}\label{CE10}
&\d\bm{C}_\z^\text{grav}+\z\.\bm{E}_{ab}^\text{grav}\d g^{ab}\\
&=d[\z\.\bm{\Q}^\text{grav}(g,\d g)-\d \bm{Q}_\z^\text{grav}]+\bm{\w}^\text{grav}(g, \d g, \math{L}_\z g)
\end{aligned}\ea
where
\ba
\w^\text{grav}(g,\d_1 g,\d_2 g)=\d_1 \bm{\Q}^\text{grav}(g, \d_2 g)-\d_2 \bm{\Q}^\text{grav}(g, \d_1 g)\quad\quad
\ea
is the symplectic current $(n-1)$-form.

After replacing $\z^a$ by $\x^a$ and noting that the configuration is static such that $\math{L}_\x g_{ab}=0$, we have $\w(g,\d g,\math{L}_\x g)=0$. Then, integration of Eq. \eq{CE10} on $C$ yields
\ba\begin{aligned}\label{CE}
&\int_C\d\bm{C}_\x^\text{grav}+\int_C\x\.\bm{E}_{ab}^\text{grav} \d g^{ab}\\
&=\int_{\pd C}[\x\.\bm{\Q}^\text{grav}(g,\d g)-\d \bm{Q}_\x^\text{grav}]\,.
\end{aligned}\ea
For the first term in Eq. \eq{CE}, using Eqs. \eq{ExpC} and \eq{rH}, we have
\ba\begin{aligned}\label{eqq1}
\int_C\d\bm{C}_\x^\text{grav}=-\int_{C}\d (\c\tilde{\bm{\epsilon}}H_{uu})=-\int_{C}\d (\c\tilde{\bm{\epsilon}}\r)
\end{aligned}\ea
where we have denoted $\tilde{\bm{\epsilon}}(\l)$ to the volume element of $\S$ in the static configuration $\f(\l)$. Substituting Eq.\eq{decdg} into the second term in Eq. \eq{CE}, we have
\ba\begin{aligned}
\int_C\x\.\bm{E}_{ab}^\text{grav} \d g^{ab}&=\int_C \tilde{\bm{\epsilon}} H_{uu}\d\c+\frac{1}{2}\int_C\c \tilde{\bm{\epsilon}} H_{ab} \d h^{ab}\\
&=\int_C \tilde{\bm{\epsilon}} \r\d\c+\frac{1}{2}\int_C\c \tilde{\bm{\epsilon}} H_{ab} \d h^{ab}\,.
\end{aligned}\ea
Combining the above results, we can further obtain
\ba\begin{aligned}\label{vareq}
&\frac{1}{2}\int_C\c \tilde{\bm{\epsilon}} H_{ab} \d h^{ab}-\int_C\c \d(\tilde{\bm{\epsilon}} \r)\\
&=\int_{\pd C}[\x\.\bm{\Q}^\text{grav}(g,\d g)-\d \bm{Q}_\x^\text{grav}]\,.
\end{aligned}\ea
For the off-shell configuration $\f=\f(0)$, the first term of left side in the above equation is not equal to $p h_{ab}$.

Next, we would like to evaluate the variation of the total entropy of perfect fluid inside $C$ when the total particle number is fixed. The total entropy of the perfect fluid is given by
\ba\begin{aligned}
S=\int_C\tilde{\bm{\epsilon}} s(\r, n)\,.
\end{aligned}\ea
The variation of the total entropy yields
\ba\begin{aligned}\label{dS1}
\d S=\int_{C} \left[s\d \tilde{\bm{\epsilon}}+\left(\frac{\pd s}{\pd \r}\d \r+\frac{\pd S}{\pd n}\d n\right)\tilde{\bm{\epsilon}}\right]\,.
\end{aligned}\ea
From the local first law $ds=\c d\r-\c \m dn$, we have
\ba\begin{aligned}
\frac{\pd s}{\pd \r}=\c\,,\quad \frac{\pd s}{\pd n}=-\c \m\,.
\end{aligned}\ea
Then, Eq. \eq{dS1} becomes
\ba\begin{aligned}\label{ddS}
\d S=\int_C\left[s \d\tilde{\bm{\epsilon}}+\c\left(\d \r-\m \d n\right)\tilde{\bm{\epsilon}}\right]\,.
\end{aligned}\ea
From the assumption that the total number of particle
\ba\begin{aligned}
N(\l)=\int_C\tilde{\bm{\epsilon}}(\l) n(\l)
\end{aligned}\ea
are fixed inside $C$, we have
\ba\begin{aligned}\label{fixedn}{
\int_Cn\d\tilde{\bm{\epsilon}}=-\int_C\tilde{\bm{\epsilon}}\d n\,.}
\end{aligned}\ea
Together with the fact that $\m \c=\text{constant}$, Eq. \eq{ddS} reduces to
\ba\begin{aligned}\label{DS}
&\d S=\int_C\left[(s+\c\m n) \d\tilde{\bm{\epsilon}}+{ \tilde{\bm{\epsilon}}}\c\d \r\right]\\
&=\int_C\left[\c \d(\tilde{\bm{\epsilon}} \r)-\frac{1}{2}\tilde{\bm{\epsilon}}\c p h_{ab}\d h^{ab}\right]\\
&=\int_C\tilde{\bm{\epsilon}}\frac{\c}{2} (\hat{H}_{ab}-p h_{ab})\d h^{ab}+\int_{\pd C}[\d \bm{Q}_\x^\text{grav}-\x\.\bm{\Q}^\text{grav}(g,\d g)]\,.
\end{aligned}\ea
In the second step, we have used the Gibbs-Duhem relation in Eq. \eq{firstlaw} and $\d \tilde{\bm{\epsilon}}=-(1/2)\tilde{\bm{\epsilon}} h_{ab}\d h^{ab}$. In the last step, we have used the variational identity in Eq. \eq{vareq}. The second part of the right side of Eq. \eq{DS} is only a boundary quantity on $\pd C$. Indeed, this boundary term corresponds to the variation of the quasi-local conserved charge corresponding to the static Killing vector $\x^a$ in the enclosed region $C$, which is defined as
\ba
Q(\x)=\int_{\pd C}\left[ \D \bm{Q}_\x^\text{grav}-\x\.\int^{\l}_{\l_0}d\l\bm{\Q}^\text{grav}(g(\l),g'(\l))\right]
\ea
with
\ba
\D \bm{Q}_\x^\text{grav}=\bm{Q}_\x^\text{grav}(g(\l))-\bm{Q}_\x^\text{grav}(g(\l_0))\,.
\ea
in which $\f(\l_0)$ of the one-parameter family $\f(\l)$ is a vacuum solution in the diffeomorphism-covariant purely gravitational theory \cite{Kim:2013zha}.  Then, throwing away this term amounts to set the variation of the quasi-local conserved charge vanishes, i.e.,
\ba\begin{aligned}
\d Q(\x)=\int_{\pd C}[\d \bm{Q}_\x^\text{grav}-\x\.\bm{\Q}^\text{grav}(g,\d g)]=0\,.
\end{aligned}\ea
For the Einstein gravity in the static spherically symmetric spacetime \cite{Gao} with the line element
\ba
ds^2=g_{tt}dt^2+\left(1-\frac{2m(r)}{r}\right)^{-1}dr^2+r^2d\W^2\,.
\ea
Let $C$ is the compact region $r\leq R$. With a straightforward calculation, $\d Q(\x)=0$ implies that the total mass $M=m(R)$ within $R$ is fixed on the boundary $r=R$. The boundary condition to fix the quasi-local conserved charge $Q(\x)$ is dependent on the explicit theories considered. For instance, in Einstein gravity or Lovelock gravity, it is to fix the induced metric and its derivative on the boundary $\pd C$ \cite{Gao,Lovelock2}; in $f(R)$ gravity, we need to fix induced metric as well as scalar curvature $R$ and its derivative on the boundary $\pd C$ \cite{Fang:2015pcw}.

For a thermodynamic system by usual matters, the entropy principle is satisfied when the system is isolated. However, for self-gravitating cases, the phrase ``isolated system'' becomes more ambiguous since the gravitational theory is diffeomorphism invariant. For a quasi-local system, i.e., $C$ is a finite compact region, the boundary condition of the isolated system should be quasi-locally imposed. By analogy to the usual cases, we should impose a boundary condition such that the variation inside the compact region $C$ would not affect the dynamics outside $C$, i.e., the variation of the spacetime will not affect the on-shell solution without the fluid outside the region $C$. That is to say, for any element of the one-parameter, their geometries outside $C$ only differ by a diffeomorphism. Under the above condition, using the on-shell variational identity \eq{CE10} outside $C$, it is easy to get
\ba\begin{aligned}
\d Q(\x)&=\int_{\pd C}[\d \bm{Q}_\x^\text{grav}-\x\.\bm{\Q}^\text{grav}(g,\d g)]\\
&=\int_{\inf}[\d \bm{Q}_\x^\text{grav}-\x\.\bm{\Q}^\text{grav}(g,\d g)]\,,
\end{aligned}\ea
where ``$\inf$'' denotes a $(n-2)$-sphere at asymptotical infinity. If the spacetime is asymptotically flat, $Q(\x)$ can be regarded as the mass $M$ of the spacetime. Since we assume that the variation inside the isolated region will not affect the on-shell geometry outside, it is natural to impose the condition such that the total mass of the spacetime fix under the variation, i.e., we have $\d Q(\x)=0$. Then, the second part of the right side of Eq. \eq{DS} vanishes and we have
\ba\begin{aligned}\label{DSres}
\d S=\frac{1}{2}\int_C\tilde{\bm{\epsilon}}\c(\hat{H}_{ab}-p h_{ab})\d h^{ab}\,.
\end{aligned}\ea
From the above result, we show that for the off-shell static configuration $\hat{H}_{ab}\neq p h_{ab}$, the variation of the total entropy is nonvanishing. In other words, The extrema of the total entropy of perfect fluid inside an isolated region $C$ for fixed total particle number demands that the static configuration is an on-shell solution. This completes the proof of \textbf{Theorem} of the entropy principle.

\section{Conclusion}\label{conc}

Our proof shows the equivalence of the extrema of the total entropy of perfect fluid in the off/on-shell static configurations inside a compact region $C$ and the dynamical equations of the static self-gravitating perfect fluid for a general diffeomorphism-covariant gravitational theory, although they are derived from two different and independent principles. The result is universal and suitable for any diffeomorphism-covariant purely gravitational theories only imposing the static condition of the spacetime. Our work provides strong evidence to show that as two independent basic principles, the ordinary thermodynamics and dynamics in the gravitational theories are consistent.

{  It's worth noting that our result is only valid for the purely gravitational theories minimally coupled to the self-gravitating perfect fluid. Is it still possible to extend  the entropy principle to some more general diffeomorphism-covariant theories for example when there is non-minimal coupling interaction between matter and gravity? For these cases, we first need to derive the concrete expressions of Neother charge $\bm{Q}_\x$ and constraint $\bm{C}_\x$, and discuss the interactions between the self-gravitating fluid and these non-minimally coupling matters. This is an interesting question and needs careful investigation in the future.
}

\section*{Acknowledgement}
Fang was supported by National Natural Science Foundation of China (NSFC) with Grants No. 11705053 and 12035005. Gao was supported by National Natural Science Foundation of China (NSFC) with Grants No. 11775022 and 11873044.

~\\

\end{document}